\documentclass[12pt]{article}

\usepackage{amsfonts,amstext,}

\usepackage{alltt}
\usepackage{calc}

\usepackage{graphicx} 
\usepackage{bm} 
\usepackage{amsmath}
\usepackage{amssymb}
\usepackage{arydshln}
\usepackage{color}

 \usepackage[top=1.0in, bottom=.750in, left=1in, right=1in]{geometry}

\newcommand{\rmd}{{\rm d}}
\newcommand{\rmi}{{\rm i}}
\newcommand{\rme}{{\rm e}}

\begin{document}

\title{Unfreezing Casimir invariants: singular perturbations giving rise to forbidden instabilities}

\author{Z. Yoshida$^1$\footnote{email: yoshida@ppl.k.u-tokyo.ac.jp} \  and \ P.J. Morrison$^2$\footnote{email: morrison@physics.utexas.edu}\\
[8pt]
$^1$ {\it Graduate School of Frontier Sciences}
\\
{\it The University of Tokyo, Kashiwa}
\\
{\it Chiba 277-8561, Japan}
\\[4pt]$^2${\it  Department of Physics and Institute for Fusion Studies} \\
{\it The University of Texas at Austin}\\
 {\it Austin, TX 78712-1060, USA}}

\maketitle

\baselineskip 20 pt

\begin{abstract}

The infinite-dimensional mechanics of fluids and plasmas  can be formulated as  ``noncanonical''
Hamiltonian systems on a phase space of Eulerian variables.
% The Poisson operator has a nontrivial kernel which 
% imposes topological constraints on dynamics.
Singularities of the Poisson bracket operator produce singular Casimir elements that
foliate the phase space, imposing topological constraints on the dynamics.
Here we proffer a physical interpretation of Casimir elements as \emph{adiabatic invariants}
---upon coarse graining microscopic angle variables, we obtain a macroscopic hierarchy on which 
the separated action variables become adiabatic invariants.  
On reflection, a Casimir element may be \emph{unfrozen} by recovering a corresponding angle variable;
such an increase in the number of degrees of freedom is, then, formulated as a \emph{singular perturbation}.
As an example, we propose a canonization of the resonant-singularity of the Poisson bracket operator of the 
 linearized magnetohydrodynamics equations, by which the ideal obstacle 
(resonant Casimir element) constraining the dynamics is unfrozen, giving rise to
a tearing-mode instability.
% The idea is based on the newly formulated ``minimum canonization''
% of noncanonical Hamiltonian system by translating Casimir invariants as adiabatic invariants.

\end{abstract}

%%%%%%%%%%%%%%%%%%%%%%%%
%%%%%%%%%%%%%%%%%%%%%%%%
%%%%%%%%%%%%%%%%%%%%%%%%
%%%%%%%%%%%%%%%%%%%%%%
%%%%%%%%%%%%%%%%%%%%%%%%%
%%%%%%%%%%%%%%%%%%%%%%%%%

\section{Introduction}
\label{sec:Introduction}

Whereas  canonical Hamiltonian mechanics is described by a Poisson bracket operator (field tensor)
that has a full rank on a symplectic manifold,  general \emph{noncanonical} Hamiltonian mechanics is endowed with a Poisson bracket operator (henceforth Poisson operator) 
that may have a nontrivial kernel; the corresponding Poisson manifold
may then be split into some local symplectic leaves (Lie-Darboux theorem).  
A Casimir element foliates the Poisson manifold
(with the gradient of a Casimir element belonging to the kernel of the Poisson operator).\footnote{
A Casimir element $C$ is a member of the \emph{center}
 of the Poisson algebra,
i.e., $[C, G]=0$ for all $G$.
For  finite-dimensional systems, one may regard phase space as the dual of the Lie algebra and  Casimir leaves as coadjoint orbits.
Unfortunately, in infinite-dimensions, there are functional analysis challenges that limit this interpretation (see, e.g., \cite{khesin}).
Note,   a Casimir leaf is not necessarily a symplectic leaf, because the kernel of
a Poisson operator may not be fully integrable when the Poisson operator has singularities
(see, e.g., \cite{narayanan,YMD}.} 
Consequently, an  orbit is constrained to a leaf (level set) of a Casimir element,
i.e.,  a Casimir element is a constant of motion.
The constancy of a Casimir element is independent of the Hamiltonian
(whereas a usual constant of motion pertains to some symmetry of a Hamiltonian), and 
it is due to a singularity of the Poisson operator.
Here we proffer an interpretation:  ``a Casimir element is an \emph{adiabatic invariant} that
is separated from a microscopic angle variable by coarse graining''
---a Casimir leaf is then a \emph{macroscopic hierarchy}.
On reflection, a Casimir invariant may be \emph{unfrozen} by recovering a corresponding angle variable.
Such an increase in the number of  degrees of freedom is, then, formulated as a \emph{singular perturbation}
(cf.\,\cite{Yoshida_Springer}).

We will cast the theory of \emph{tearing-mode instabilities} into a perspective of  an unfrozen Casimir leaf.  The dynamics of an ideal plasma obeys topological constraints on the co-moving magnetic field,\footnote{
The core of the topological constraint is epitomized by Kelvin's circulation law: 
in a barotropic fluid, the Lie derivative of fluid momentum is an exact differential 1-form;
thus,  the circulation (co-moving loop-integral of the fluid momentum, or the surface-integral of the vorticity) is conserved.
In a plasma (charged fluid), the momentum combines with the electromagnetic potential (gauge field),
and the corresponding \emph{canonical vorticity} is the combination of the fluid vorticity and magnetic field.
In the magnetohydrodynamic model, the canonical vorticity of the electron fluid is
approximated by the magnetic field, neglecting the electron mass.  
The circulation law for the magnetic field implies that the magnetic flux on each fluid element is conserved, the so-called Alfv\'en's law.} 
which prevents the onset of various potential instabilities. 
A ``tearing mode'' is a mode of deformation that brings about a change in the topology of
magnetic surfaces (the integral surfaces of magnetic field lines), 
which is, therefore, forbidden to occur in the ideal dynamics\,\cite{FKR63,Furth63,White83}.
In\,\cite{YD2012}, 
% we have identified the ``helical flux Casimir'' that poses an ideal obstacle on the growth of tearing mode;
a tearing mode is formulated as an equilibrium point on a helical-flux-Casimir leaf.
As long as the helical-flux Casimir is constrained, the tearing-mode cannot grow.
We will formulate a ``perturbation'' that allows the system to change the
helical flux, as well as absorb (dissipate) the energy.
An unstable tearing mode has  negative-energy, with respect to a fiducial ``helical''
bifurcated equilibrium; hence, 
a tearing-mode instability is a process of moving across Casimir leaves toward a lower-energy
equilibrium point with a different (helical) topology.

In the next section, we begin by reviewing some aspects of the basic framework of Hamiltonian mechanics.  
In Sec.\,\ref{sec:adiabatic}, we then consider an example of magnetized particles, in order  to establish a connection between adiabatic invariants and Casimir elements.
In Sec.\,\ref{sec:canonicaliztion}, we will  formulate a systematic method of \emph{canonization} by adding ``angle variables'' that result in the unfreezing of Casimir elements.
As is now well known, the  infinite-dimensional mechanics of a plasma can be formulated as a noncanonical
Hamiltonian system on a phase space of Eulerian variables (see e.g.\, \cite{morrison98}).
After a short review of the Hamiltonian formalism of magnetohydrodynamics (MHD)
and its application to the tearing-mode theory (Sec.\,\ref{subsec:Beltrami}),
we will  formulate a (formal) singular perturbation that gives rise to a tearing-mode
instability, and finally discuss its physical implications (Sec.\,\ref{subsec:tearing-mode}).
 
%%%%%%%%%%%%%%%%%%%%%%%%%
%%%%%%%%%%%%%%%%%%%%%%%%%
%%%%%%%%%%%%%%%%%%%%%%%%%
%%%%%%%%%%%%%%%%%%%
%%%%%%%%%%%%%%%%%%%%%%%%%%
%%%%%%%%%%%%%%%%%%%%%%%%%%
%%%%%%%%%%%%%%%%%%%%%%%%%%

\section{Preliminaries: noncanonical Hamiltonian systems and Casimir invariants}
\label{sec:preliminaries}

%%%%%%%%%%%%%%%%%%%%%%%%%%%%%%%%%%%%%%%%%%%%%%%%%%%%%%%%%%%%%%%%%%%%%%%%%%%%%%%%%%%%%%%%%%%%%%
% \subsection{Poisson operators and Casimir invariants}
% \label{subsec:Possion_Manifold}

We denote by $\bm{z}=(q^1, \cdots, q^m, p^1,\cdots,p^m)$ the \emph{state vector}, a point in an
affine space $X=\mathbb{R}^{2m}$ (to be called \emph{phase space}).\footnote{
Usually, phase space is identified as a cotangent bundle $T^*M$ of a
smooth manifold $M$ of dimension $m$, on which 
a symplectic 2-form $\omega =(1/2) J_{c,k\ell} dz^k\wedge dz^\ell$
(the vorticity of a canonical 1-form) defines symplectic geometry.}
A canonical Hamiltonian system is endowed with a \emph{Hamiltonian}
$H(\bm{z})$ (a real function on the phase space $X$) and a 
$2m\times 2m$ antisymmetric regular matrix
\[
J_c := 
\left(
\begin{array}{cc}
0_m & I_m \\
-I_m & 0_m 
\end{array}
\right),
\]
where $I_m$ and $0_m$ are the $m$-dimensional identity and nullity, respectively.  
(In what follows, we will write just $I$ or $0$ without specifying the dimension,
especially when we consider an infinite-dimensional space.) 
We call $J_c$ the  \emph{canonical Poisson operator} (matrix).
The equations of motion (Hamilton's equations) are written as
\begin{equation}
\frac{\rmd}{\rmd t} \bm{z} = J_c \partial_{\bm{z}} H(\bm{z}) ,  %\equiv P(\bm{z}) H(\bm{z}).
\label{Hamilton_eq_canonical}
\end{equation}
whence an equilibrium point is seen to satisfy $\partial_{\bm{z}}H(\bm{z})=0$. 
Defining a Poisson bracket by
\[
[ a, b ] :=  (\partial_{z_i} a)J_{ij}(\partial_{z_j} b), 
\]
 the rate of change of an observable $f(\bm{z})$ is determined by
\[
\frac{\rmd}{\rmd t} f = [ f, H ].
\]

One may generalize the  Poisson operator $J$ to be a function
$J(\bm{z})$ of an arbitrary dimension $n\times n$
(here we assume a finite $n$, while we will consider infinite-dimensional systems later).
A \emph{noncanonical} Hamiltonian system allows $J(\bm{z})$ to be singular,
i.e.,  $\textrm{Rank}\,J(\bm{z})$ may be less than $n$ and can change as a function of $\bm{z}$
(while the corresponding Poisson bracket must satisfy Jacobi's identity).
The equations of motion are, then, 
\begin{equation}
\frac{\rmd}{\rmd t} \bm{z} = J(\bm{z}) \partial_{\bm{z}} H(\bm{z}).  % =: P(\bm{z}) H(\bm{z}).
\label{Hamilton_eq_1}
\end{equation}
% Here $P(\bm{z})$ is a first-order partial differential operator.
% The fixed points may not be only the critical points of the Hamiltonian $H(\bm{z})$.
A \emph{Casimir element} $C(\bm{z})$ is a solution to a
partial differential equation (PDE)
\begin{equation}
J(\bm{z}) \partial_{\bm{z}} C(\bm{z}) =0,
\label{Casmir-1}
\end{equation}
which implies that $[C, F]=0$ for every $F$.
Therefore, $C$ is a constant of motion ($dC/dt=[C,H]=0$ for any Hamiltonian $H$).

Obviously, if $\mathrm{Rank}\,J(\bm{z})=n$ (the dimension of the phase space), 
(\ref{Casmir-1}) has only the trivial solution ($C=$ constant).
If the dimension $\nu$ of $\textrm{Ker}(J(\bm{z}))$ does not change, 
the solution of (\ref{Casmir-1}) may be
constructed by ``integrating'' the elements of $\textrm{Ker}(J(\bm{z}))$
---then the Casimir leaves are symplectic manifolds. 
This expectation turns out to be true provided the Poisson bracket satisfies the 
Jacobi  identity and $m-\nu$ is an even number (Lie-Darboux theorem).
However, the point where the rank of $J(\bm{z})$ changes is a  singularity of the 
PDE (\ref{Casmir-1})\,\cite{morrison98}, 
from which singular Casimir elements are generated\,\cite{YMD}.
 
% \subsection{Energy-Casimir function} %-----------------------------------
% \label{subsec:energy-Casimir}

When we have a Casimir element $C(\bm{z})$ in a noncanonical Hamiltonian system,
a transformation of the Hamiltonian $H_(\bm{z})$ such as
\begin{equation}
H(\bm{Z}) \mapsto {H}_{{\mu}}(\bm{z})
= {H}(\bm{z}) - \mu {C} (\bm{z})
% \quad (\mu\in\mathbb{R})
\label{Hamiltonian-system-3}
\end{equation}
(with an arbitrary real constant $\mu$) does not change the dynamics.
In fact, Hamilton's equations  (\ref{Hamilton_eq_1}) are  invariant under this transformation.
% \begin{equation}
% \frac{d}{dt}\bm{z}
% = J \partial_{\bm{z}} {H}_{{\mu}}(\bm{z})
% = J \partial_{\bm{z}} {H}(\bm{z}) .
% \label{Hamiltonian-system-2'}
% \end{equation}
We call the transformed Hamiltonian 
${H}_{{\mu}}(\bm{z})$ an  \emph{energy-Casimir} function\,
\cite{lausanne,energy-casimir,morrison98,Arnold-Khesin}.

Interpreting the parameter $\mu$ as a Lagrange multiplier of the equilibrium variational principle,
${H}_{{\mu}}(\bm{z})$ is the effective Hamiltonian with the constraint that restricts the
Casimir element $C(\bm{z})$
to be a given value (since $C(\bm{z})$ is a constant of motion, its value is fixed by its
initial value).
As we will see in some examples, Hamiltonians are rather simple, often being ``norms'' on  the phase space.
However, an energy-Casimir functional may have a nontrivial structure.
Geometrically, ${H}_{{\mu}}(\bm{z})$ is the distribution of $H(\bm{z})$ on a 
Casimir leaf (hyper-surface of $C(\bm{z})=$ constant).  If Casimir leaves are
distorted with respect to the energy norm, the effective Hamiltonian 
may have a complex distribution on the leaf, which is, in fact, the origin of
various interesting structures in noncanonical Hamiltonian systems.

%%%%%%%%%%%%%%%%%%%%%%%%
%%%%%%%%%%%%%%%%%%%%%%%%
%%%%%%%%%%%%%%%%%%%%%%%%
%%%%%%%%%%%%%%%%%%%%%%
%%%%%%%%%%%%%%%%%%%%%%%%%
%%%%%%%%%%%%%%%%%%%%%%%%%
%%%%%%%%%%%%%%%%%%%%%%%%%% 
\section{Foliation by adiabatic invariants}
\label{sec:adiabatic}
Here we study an example of noncanonical Hamiltonian mechanics
(and creation of interesting structures on Casimir leaves) in which
Casimir elements originate from adiabatic invariants.

The Hamiltonian of a charged particle is the sum of the
kinetic energy and the potential energy:
$H = m v^2/2 + q \phi$,
where $\bm{v} := (\bm{P}-q\bm{A})/m$ is the velocity,
$\bm{P}$ is the canonical momentum, $(\phi,\bm{A})$ is the electromagnetic 4-potential,
$m$ is the particle mass, and $q$ is the charge.
Needless to say, a magnetic field does not change the value of energy, and 
the standard Boltzmann distribution function is independent to the magnetic field.
However, in the vicinity of a dipole magnetic field rooted in a star or  planet,
for example, we often find a plasma clump with a rather steep density gradient.
In such a situation,   so-called \emph{inward diffusion} drives charged particles
toward the inner higher-density region, which 
is seemingly opposite to the natural direction of diffusion (normally, diffusion is a process of
flattening distributions of physical quantities). 
Creation of such a macroscopic structure can be explained only by delineating a 
fundamental difference between  a macroscopic hierarchy and  basic microscopic mechanics.
Since the magnetic field does not cause any change in the energy of particles, there is no way to
revise the energy in the calculation of the equilibrium state.
Instead, the problem is solved by finding an appropriate ``phase space'' (or an ensemble) on which the Boltzmann distribution is achieved; 
the identification of an appropriate macroscopic phase-space is
nothing but the formulation of what we call a ``scale hierarchy''.

Magnetized particles live in an effective phase space that is \emph{foliated} by
adiabatic invariants associated with periodic motions  of particles.
Denoting by $\bm{v}_\parallel$ and $\bm{v}_\perp$ the parallel and perpendicular 
(with respect to the local magnetic field) components of the velocity,
we may write
\begin{equation}
H = \frac{m}{2} v_\perp^2 + \frac{m}{2} v_\parallel^2 + q \phi.
\label{Hamiltonian-2}
\end{equation}
The velocities are related to the mechanical momentum via  $\bm{p}:=m\bm{v}$, $\bm{p}_\parallel:=m\bm{v}_\parallel$, and $\bm{p}_\perp:=m\bm{v}_\perp$.
In a strong magnetic field, $\bm{v}_\perp$ can be decomposed into a
small-scale cyclotron motion $\bm{v}_c$ and a macroscopic guiding-center drift motion
${\bm{v}}_d$.
The periodic cyclotron motion $\bm{v}_c$ can be ``quantized''
to write $mv_c^2/2 = \mu\omega_c(\bm{x})$
in terms of the magnetic moment $\mu$
and the cyclotron frequency $\omega_c(\bm{x})$; the adiabatic invariant $\mu$ 
and the gyration phase $\vartheta_c:=\omega_c t$ constitute an action-angle pair.
% In the standard interpretation, in analogy with the Landau levels in quantum theory,
% $\omega_c$ is the \emph{energy level}
% and $\mu$ is the \emph{number} of quasi-particles
% (quantized periodic motions) at the corresponding energy level.
The macroscopic part of the perpendicular kinetic energy is expressed as
$m v_d^2 /2 = (P_\theta-q\psi)^2/(2mr^2)$,
where $P_\theta$ is the angular momentum in the $\theta$ direction
and $r$ is the radius from the geometric axis.
In terms of the canonical-variable set
$\bm{z}=(\vartheta_c,\mu,\zeta,p_\parallel,\theta,P_\theta)$,
the Hamiltonian of the guiding center (or, the quasi-particle) becomes %\,\cite{canonical_variables}
\begin{equation}
H_c = \mu \omega_c + \frac{1}{2m} p_\parallel^2
+ \frac{1}{2m}\frac{\left( P_\theta-q\psi\right)^2}{r^2}
+ q\phi .
\label{Hamiltonian-3}
\end{equation}
Note that the energy of the cyclotron motion has been quantized in term of the frequency $\omega_c(\bm{x})$ and the action $\mu$; the gyro-phase $\vartheta_c$ has been coarse grained (integrated to yield $2\pi$).

Now,  we formulate the ``macroscopic hierarchy'' 
on which charged particles create a thermal equilibrium.
The adiabatic invariance of the magnetic moment $\mu$
imposes a \emph{topological constraint} on the motion of particles;
it is this constraint that is the root-cause of a macroscopic hierarchy and of structure formation. 
% Mathematically, the \emph{scale hierarchy} is equivalent to a foliation of the phase space.
% To explain how the scale hierarchy is formulated, we start
% by the general (micro-macro total) formulation, and then
% separate the microscopic action-angle pair $\mu$-$\vartheta_c$;
% the \emph{macroscopic phase space} is the remaining sub-manifold 
% immersed in the general phase space, which we 
% delineate as a leaf of the foliation in terms of a \emph{Casimir invariant}
% ---if there is a nontrivial function $C$
% satisfying $\{ G, C \}=0$ for every $G$,
% we say that the Poisson bracket $\{~,~\}$ is \emph{noncanonical},
% and call $C$ a \emph{Casimir invariant}; see Sec.\,\ref{sec:noncanonical}.
The Poisson operator on the total (microscopic) phase space, spanned by the canonical variables
$\bm{z}=(\vartheta_c,\mu,\zeta,p_\parallel,\theta,P_\theta)$, is %defined as
% \[
% \{ F,G \} := ( \partial_{\bm{z}} F, \mathcal{J} \partial_{\bm{z}} G ),
% \]
% where $( \bm{u}, \bm{v} ) := \int u_j v^j {d}^6z$ is the inner-product and
% ${J}$ is 
a canonical symplectic matrix:
\begin{equation}
J :=\left( \begin{array}{ccc}
J_c & 0 & 0 \\
0 & J_c & 0 \\
0 & 0 & J_c  \end{array} \right),
\quad 
J_c := \left( \begin{array}{cc}
0 & 1 \\
-1& 0    \end{array} \right).
\label{symplectic-canonical}
\end{equation}
The equations of motion for the Hamiltonian $H_c$ are  written as ${\rmd} z^j/\rmd t = [ z^j,H_c]$.
Notice that the quantization of the cyclotron motion in $H_c$
suppresses change in $\mu$.
% Liouville's theorem determines the invariant measure $\rmd^6z$, by which
% we obtain the Boltzmann distribution (\ref{Boltzmann-2}).

To extract the macroscopic hierarchy, we ``separate out''
the microscopic variables $(\vartheta_c,\mu)$ 
by modifying the Poisson operator as follows:
\begin{equation}
{J}_{nc} :=\left( \begin{array}{ccc}
0 & 0 & 0 \\
0 & J_c & 0 \\
0 & 0 & J_c \end{array} \right).
\label{symplectic-noncanonical}
\end{equation}
The Poisson bracket
$[ F, G ]_{nc} := \langle\partial_{\bm{z}}F, {J}_{nc}\partial_{\bm{z}}G\rangle$
determines the kinematics on the macroscopic hierarchy;
The corresponding kinetic equation
$\partial_t f + [ H_c, f ]_{nc} =0$ reproduces the familiar drift-kinetic equation.
The kernel of ${J}_{nc}$ makes the Poisson bracket $[\, ,\,]_{nc}$ \emph{noncanonical}\,\cite{morrison98}.
Evidently, $\mu$ is a Casimir element
(more generally $C=g(\mu)$ with $g$ being any smooth function).
The level-set of $\mu$, a leaf of the Casimir foliation, identifies
what we may call the \emph{macroscopic hierarchy}.

By applying Liouville's theorem to the Poisson bracket $[\, ,\,]_{nc}$, 
the invariant measure on the macroscopic hierarchy 
is $\rmd^4z=\rmd^6z/(2\pi \rmd\mu)$, the the total phase-space measure modulo the
microscopic measure.
The most probable state (statistical equilibrium) on the macroscopic ensemble
maximizes the entropy % with respect to this invariant measure.
% The variational principle is set up following the standard procedure
% ---immersing the macroscopic hierarchy into the general phase space, 
% and incorporating the constraints through the Lagrange multipliers:
$S=-\int f\log f\,\rmd^6z$ 
for a given particle number $N=\int f \rmd^6z$, a quasi-particle number $M=\int \mu f \rmd^6z$,
and an energy $E=\int H_c f \rmd^6z$.
Then, the distribution function is
\begin{equation}
f = f_\alpha := Z^{-1} {e}^{ -(\beta H_c + \alpha\mu)},
\label{modified_Boltzmann}
\end{equation}
where $\alpha$, $\beta$, and $\log Z-1$ are, respectively the Lagrange multipliers on
$M$, $E$, and $N$.
In this  \emph{grand-canonical distribution function}, 
$\alpha/\beta$ is the chemical potential associated with the quasi-particles.\footnote{
We can also derive (\ref{modified_Boltzmann}) by 
an \emph{energy-Casimir function}.
With a Casimir element $\mu$, we can transform the Hamiltonian as $H_c \mapsto H_\alpha
:= H_c + \alpha \mu$ ($\alpha$ is an arbitrary constant)
without changing the macroscopic dynamic.
The Boltzmann distribution with respect to $H_\alpha$ is equivalent to (\ref{modified_Boltzmann}). This equivalency was discussed in greater generality in \cite{morrison87}.}
The factor ${e}^{-\alpha\mu}$ in $f_\alpha$
yields a direct $\omega_c$ dependence of the coordinate-space density:
\begin{equation}
\rho =\int f_\alpha \, \frac{2\pi\omega_c }{m} \rmd\mu \rmd v_d \rmd v_\parallel
\propto \frac{\omega_c(\bm{x})}{\beta\omega_c(\bm{x})+\alpha } ,
\label{density-3}
\end{equation}
which demonstrates the creation of a density clump near the dipole magnetic field\,\cite{RT-1_PPCF2013}.
% which may be compared with the density (\ref{density-2}) evaluated for the Boltzmann distribution
% ($\phi=0$ assuming charge neutrality).
% Notice that the Jacobian $(2\pi\omega_c /m)\rmd\mu$ multiplying the macroscopic 
% measure $\rmd^4z$ reflects the distortion of the macroscopic phase space (Casimir leaf)
% caused by the magnetic field.
% Figure\,\ref{fig:density} shows the density distribution and the magnetic field lines.

%%%%%%%%%%
%%%%%%%%%%%%%%%%%%%%%%%%%%%%
%%%%%%%%%%%%%%%%%%%%%%%%%%%%%
%%%%%%%%%%%%%%%%%%%%%
%%%%%%%%%%%%%%%%
%%%%%%%%%%%%%%%%%%%%%%%%%%%%
%%%%%%%%%%%%%%%%%%%%%%%%%%%%%
%%%%%%%%%%%%%%%%%%%%%%%%%%%%%%%%%%%%%
\section{Canonization atop Casimir leaves}
\label{sec:canonicaliztion}

The aim of this section is to formulate a systematic method of
 ``canonization'' of a noncanonical system
by embedding the system into  a higher-dimension phase space;
Casimir elements become  ``adiabatic invariants'' associated with a symmetry 
(at a macro-scale hierarchy) of a Hamiltonian.

%---------------------------------------------------------------------------------
\subsection{Extension of the phase space and canonization}

Let $J$ be a Poisson  matrix on an $n$-dimensional 
phase space $X=\mathbb{R}^n$ parameterized by $\bm{z} = (z_1, \cdots, z_n)$.
We assume that $\textrm{Ker}(J)$ has a dimension $\nu$ and
$n-\nu$ is an even number.  We also assume that $\textrm{Ker}(J)$ is spanned  by
Casimir invariants $C_1,\cdots,C_\nu$, i.e.
\begin{equation}
\textrm{Ker}(J) = \{ \nabla C_1, \cdots, \nabla C_\nu \} .
\label{foliation}
\end{equation}
Our mission is to find the ``minimum'' extension of the phase space and
a \emph{canonical} Poisson  matrix $\tilde{J}$ by which the Casimir invariants
are re-interpreted as adiabatic invariants ---an appropriate perturbation of the
Hamiltonian will then give a near-integrable system in the vicinity of the
original Casimir leaves.
The target phase space must be of dimension $\tilde{n}:=n+\nu$
(even number) consisting of $z_1,\cdots,z_n$ and additional
$\vartheta_1,\cdots,\vartheta_\nu$.  

Before formulating such a minimum system, we note that we may formally produce a
``larger'' system; 
the simplest method of extension and canonization is to double the phase space:
let $\bm{z}_{\times2} := (z_1,\cdots,z_n,\chi_1, \cdots, \chi_n)$,
and 
\begin{equation}
J_{\times2} := 
\left( \begin{array}{c:c}
J & L(\bm{\chi}) \\
\hdashline
-^t L(\bm{\chi}) & 0
\end{array} \right),
\label{X2}
\end{equation}
where $L(\bm{\chi})$ is a certain regular $n\times n$ matrix.
To satisfy the Jacobi  identity,
$L(\bm{\chi})$ must satisfy the Maurer-Cartan equation; see Eq.~(292) of 
\cite{morrison98}.
% We note that a more general $L(\bm{z},\bm{\chi})$ is, in general, difficult to satisfy Jacobi's identity.

%---------------------------------------------------------------------------------
\subsection{``Minimum'' canonization invoking Casimir invariants}

It is generally difficult to reduce $J_{\times2}$ to $\tilde{J}$ of dimension
$\tilde{n}\times\tilde{n}$;
to separate $2n-\tilde{n}$ variables from $\bm{z}_{\times2}$, these 
variables and the remaining %$(z_1,\cdots,z_n,\vartheta_1,\cdots,\vartheta_\nu)$ 
$\tilde{n}$ variables must be independent, implying  a ``separation of variables.''

% since the additional elements to $J$ will be functions of $\bm{z}$, Jacobi's identity is not obvious.

Our strategy is to  use  the Casimir foliation (\ref{foliation}) of the
phase space.  We first canonize  $J$ on $X/\textrm{Ker}(J)$.
Let
\[
\bm{z}'= (\zeta_1,\cdots,\zeta_{n-\nu},C_1,\cdots,C_\nu) \in \mathbb{R}^n,
\]
by which $J$ is transformed into a Darboux standard form:
\begin{equation}
J' =
\left( \begin{array}{ccc:c}
~   & ~   &  ~   &  ~   \\
 ~  &J_c  & ~    &  ~   \\
 ~  &   ~ & ~    &  ~   \\
  \hdashline
 ~  &  ~  &  ~   & 0_\nu
\end{array} \right) ,
% \left( \begin{array}{c:ccc}
% 0_\nu & 0  & \cdots &0 \\
% \hdashline
% 0 & J_c & 0  & 0\\
% \vdots & 0   & \ddots & 0 \\
% 0  &  0  & 0  & J_c
% \end{array} \right) ,
\label{J-canonical}
\end{equation}
We can extend $J'$ to an $\tilde{n}\times\tilde{n}$ canonical matrix such that
\begin{equation}
{J}_{ex} =
\left( \begin{array}{ccc:cc}
~   & ~   &  ~   &  ~ & ~  \\
 ~  &J_c  & ~    &  ~ & ~  \\
 ~  &   ~ & ~    &  ~ & ~  \\
  \hdashline
 ~  &  ~  &  ~   & 0_\nu &-I_\nu \\
 ~  &  ~  &  ~   &I_\nu &0_\nu 
\end{array} \right) .
% \left( \begin{array}{c:c:ccc}
% 0_\nu  & I_\nu & 0  & \cdots &0 \\
% \hdashline
% -I_\nu & 0_\nu & 0  & \cdots &0 \\
% \hdashline
% 0      & 0 & J_c & 0  & 0\\
% \vdots & \vdots & 0   & \ddots & 0 \\
% 0      & 0  &  0  & 0  & J_c
% \end{array} \right) ,
\label{J-canonical-extend}
\end{equation}
% where $I_\nu$ is $\nu\times\nu$ identity.
The corresponding variables are denoted by
\[
{\bm{z}}_{ex} 
% \bm{z}'\oplus\bm{\vartheta} % \bm{C}\oplus\bm{\zeta}
=(\zeta_1,\cdots,\zeta_{n-\nu},C_1,\cdots,C_\nu,\vartheta_1,\cdots,\vartheta_\nu) \in \mathbb{R}^{\tilde{n}} .
\]
% Transforming $\tilde{\bm{z}}' \mapsto \tilde{\bm{z}}=\bm{\vartheta}\oplus\bm{z}$, we obtain
% \begin{equation}
% \tilde{J} =
% \left( \begin{array}{c:c}
% 0_\nu  & X  \\
% \hdashline
% -^tX & J 
% \end{array} \right) ,
% \label{J-canonical-extend2}
% \end{equation}
% where
% (denoting the Poisson bracket on $\tilde{\bm{z}}'$ by $[a, b]$)
% \begin{equation}
% X_{ij} := [\vartheta_i, z_j ] 
% \quad (i=1,\cdots,\nu,~j=1,\cdots,n).
% \label{J-canonical-extend3}
% \end{equation}

An interesting property of this extended, canonized Poisson  matrix ${J}_{ex}$ is that
the elements are independent of  the additional variables $\bm{\vartheta}$,
which is in marked contrast to the simple extension $J_{\times2}$ defined in (\ref{X2}).

%%%%%%%%%%%%%%%%%%%%%%%%%%%%%%%%%%%%%%
%%%%%%%%%%%%%%%%%%%%%%%%%%%%
%%%%%%%%%%%%%%%%%%%%%%%%%%%%
%%%%%%%%%%%%%%%%%%%%%%%%%%%%
%%%%%%%%%%%%%%%%%%%%%%%%%%%%
%%%%%%%%%%%%%%%%%%%%%%%%%%%%%%%%%%%%%%
%%%%%%%%%%%%%%%%%%%%%%%%%%%%%
%%%%%%%%%%%%%%%%%%%%%%%%%%%%%
%%%%%%%%%%%%%%%%%%%%%%%%%%%%%%%%%%%%
\section{Application to  tearing-mode theory}
\label{sec:tearin-mode}

In this section, we put the method of unfreezing Casimir elements to the test by
studying the tearing-mode instability from  the perspective of the noncanonical Hamiltonian formalism.
The system is of infinite dimension, hence the formulation needs an appropriate functional analytical setting.
Here we invoke a simple incompressible ideal MHD model.

%%%%%%%%%%%%%%%%%%%%%%%%%%%%%%%%%%%
%%%%%%%%%%%%%%%%%%%%%%%%%%%%%%
%%%%%%%%%%%%%%%%%%%%%%%%%%%%%
\subsection{Helicity and Beltrami equilibria} 
\label{subsec:Beltrami}

%---------------------------------------------------------------------------------
\subsubsection{Magnetohydrodynamics (MHD) system}

Let $\bm{V}$ and $\bm{B}$ denote the fluid velocity and magnetic field of a plasma.
Here we consider an incompressible flow, $\nabla\cdot\bm{V}=0$, hence
both $\bm{V}$ and $\bm{B}$ are solenoidal vector fields.
The governing equations are (in the so-called Alfv\'en units)
\begin{equation}
\begin{array}{l}
{\small \partial_t \bm{V} - \bm{V}\times (\nabla\times\bm{V}) 
= -\nabla p
+(\nabla\times\bm{B})\times\bm{B},}
\\
{\small \partial_t\bm{B}=\nabla\times(\bm{V}\times\bm{B}) .}
\end{array} ,
\label{MHD}
\end{equation}
where $p$ denotes the fluid pressure.
We consider a three-dimensional bounded domain $\Omega$ surrounded by a perfectly 
conducting boundary $\partial\Omega$; the boundary conditions are
(denoting by $\bm{n}$ the normal trace onto $\partial\Omega$)
\begin{equation}
\bm{n}\cdot\bm{V}=0,
\quad
\bm{n}\cdot\bm{B}=0,
\quad (\mathrm{on}~\partial\Omega).
\label{MHD-BC}
\end{equation}
The state vector $\bm{u}=\,^t(\bm{V},\bm{B})$ belongs to the phase space
$X=L^2_\sigma(\Omega)\times L^2_\sigma(\Omega)$, where
\begin{equation}
L^2_\sigma(\Omega) := \{ \bm{u}\in L^2(\Omega);\, \nabla\cdot\bm{u}=0,\,
\bm{n}\cdot\bm{u}=0 \},
\label{L2sigma}
\end{equation}
which is a closed subspace of $L^2(\Omega)$
(we endow the Hilbert space $X$ with the standard $L^2$ inner product $\langle\bm{u},\bm{v}\rangle$ and 
the norm $\|\bm{u}\|$).
We denote by $\mathcal{P}_\sigma$ the projector onto $L^2_\sigma(\Omega)$.
Defining a Hamiltonian and a Poisson operator by
\begin{eqnarray}
{H}(\bm{u}) &:=& \frac{1}{2} \left( \|\bm{V} \|^2 + \|\bm{B}\|^2 \right).
\label{MHD-Hamiltonian}
\\
{\cal J}(\bm{u}) &:=& {\small \left( \begin{array}{cc}
 -\mathcal{P}_\sigma(\nabla\times\bm{V})\times & \mathcal{P}_\sigma(\nabla\times\circ)\times\bm{B}
\\
\nabla\times\left[ \circ\times \bm{B} \right] & 0
\end{array} \right) },
\label{MHD-J}
\end{eqnarray}
the MHD system (\ref{MHD}) is cast into the following Hamiltonian form:
\begin{equation}
\partial_t \bm{u}= {\cal J}(\bm{u}) \partial_{\bm{u}} H(\bm{u}),
\label{Hamilton's_equation_general}
\end{equation}
(cf.\,\cite{MG80,morrison98,chandre}) where $\partial_{\bm{u}}$ is the 
gradient (of Lipschitz continuous functionals\,\cite{Clarke1975})  in the Hilbert 
space $X$. Here, we define ${\cal J}(\bm{u})$ on a subdomain of $C^\infty$-functions in the
phase space $X$,
which suffices to find regular equilibrium points (cf.\,\cite{YMD} for more
precise definitions).

%---------------------------------------------------------------------------------
\subsubsection{Beltrami eigenfunctions}
The Poisson operator ${\cal J}(\bm{u})$ has two independent Casimir elements
(denoting by $\bm{A}$ the vector potential of $\bm{B}$)
\begin{equation}
C_1(\bm{u}) := \frac{1}{2}\int_\Omega \bm{A}\cdot\bm{B}~\rmd^3x,
\quad
C_2(\bm{u}) := \int_\Omega \bm{V}\cdot\bm{B}~\rmd^3x,
\label{M-helicity}
\end{equation}
which, respectively, represent
the magnetic helicity and the cross helicity.
\index{helicity}
They impose topological constraints on the field lines\,\cite{Moffatt}.
The ``Beltrami equilibrium'' is an equilibrium point of the energy-Casimir functional
\index{Beltrami equilibrium}
${H}(\bm{u}) - \mu_1 C_1(\bm{u})-\mu_2 C_2(\bm{u})$.
Here we consider a subclass of equilibrium points assuming $\mu_2=0$.
Then, $\bm{V}=0$ (invoking $\mu_2\neq0$, we obtain a larger set of equilibria with a finite $\bm{V}$).
The determining equation for $\bm{B}$ is (denoting $\mu_1=\mu$)
\begin{equation}
\nabla\times\bm{B} - \mu\bm{B} = 0,
\label{beltrami-2'}
\end{equation}
which reads as an eigenvalue problem of the curl operator\,\cite{YG1990}.  
% We call (\ref{beltrami-2'}) the Beltrami equation.
The solution (to be denoted by $\bm{B}_\mu$)
is often called a \emph{Taylor relaxed state}\,\cite{JBT74,JBT86}.

While the Beltrami equation (\ref{beltrami-2'}) 
together with the homogeneous boundary conditions (\ref{MHD-BC})
are seemingly homogeneous equations, there is a ``hidden inhomogeneity'' when
$\Omega$ is multiply connected [then, the boundary conditions (\ref{MHD-BC}) are
insufficient to determine a unique solution].
To delineate the ``topological inhomogeneity'' of the Beltrami equation, we first make 
$\Omega$ into a simply connected domain $\Omega_S$ by inserting cuts $\Sigma_\ell$ across each handle of $\Omega$:
$\Omega_S:=\Omega\setminus(\cup_{\ell=1}^\nu \Sigma_\ell)$
(where $\nu$ is the \emph{genus} of $\Omega$).
The \emph{fluxes} of $\bm{B}$ are given by
(denoting by $\rmd\bm{\sigma}$ is the surface element on $\Sigma_\ell$)
$\Phi_\ell(\bm{B}) := \int_{\Sigma_\ell} \bm{B}\cdot \rmd\bm{\sigma}$,
% \label{flux}
% \end{equation}
which are constants of motion.
To separate these fixed degrees of freedom, we invoke the Hodge--Kodaira decomposition
$L^2_\sigma(\Omega) = L^2_\Sigma(\Omega) \oplus L^2_{\rm H}(\Omega) $, where
% \index[subject]{Hodge--Kodaira decomposition}
\begin{subequations}
% \label{grp}
\begin{align}
% L^2_\sigma(\Omega) &:= \{ \bm{u}\in L^2(\Omega);\, \nabla\cdot\bm{u}=0,\,
% \bm{n}\cdot\bm{u}=0 \},
% \label{L2sigma}\\
L^2_\Sigma(\Omega) &:= \{ \bm{u}\in L^2(\Omega);\, \nabla\cdot\bm{u}=0,\,
\bm{n}\cdot\bm{u}=0,\, \Phi_\ell(\bm{u})=0~(\forall\ell) \}.
\label{L2Sigma}\\
L^2_{\rm H}(\Omega) &:= \{ \bm{u}\in L^2(\Omega);\, \nabla\times\bm{u}=0,\,\nabla\cdot\bm{u}=0,\,
\bm{n}\cdot\bm{u}=0 \}.
\label{L2H}
\end{align}
\end{subequations}
The dimension of $L^2_{\rm H}(\Omega)$, the space of \emph{harmonic fields} (or \emph{cohomologies}),
is equal to the genus $\nu$ of $\Omega$.
% and $L^2_{\rm H}(\Omega)$ is spanned by gradients of 
% \emph{angle variables} $q_\ell$ ($\ell=1,\cdots,\nu$).
% We can now state the orthogonal \emph{Hodge--Kodaira decomposition}:
% \begin{equation}
% L^2_\sigma(\Omega) = L^2_\Sigma(\Omega) \oplus L^2_{\rm H}(\Omega) .
% \label{Hodge-Kodaira}
% \end{equation}
We decompose the total $\bm{B}\in L^2_\sigma(\Omega)$ into the fixed harmonic ``vacuum'' field $\bm{B}_{\rm H}\in L^2_{\rm H}(\Omega)$ (which carries the given fluxes $\Phi_1,\cdots,\Phi_\nu$) and a residual component $\bm{B}_\Sigma$ driven by currents within the plasma volume $\Omega$,
\begin{equation}
\bm{B} = \bm{B}_\Sigma + \bm{B}_{\rm H},
\quad [\bm{B}_\Sigma:={\cal P}_\Sigma \bm{B}\in L^2_\Sigma(\Omega),\, \bm{B}_{\rm H}\in L^2_{\rm H}(\Omega)],
\label{decomposition-B}
\end{equation}
where ${\cal P}_\Sigma$ denotes the orthogonal projector from $L^2(\Omega)$ onto $L^2_\Sigma(\Omega)$.

Now the Beltrami equation (\ref{beltrami-2'}) reads as an inhomogeneous equation
(denoting $\nabla\times$ by curl):
\begin{equation}
(\textrm{curl}- \mu)\bm{B}_\Sigma = \mu \bm{B}_{\rm H} ,
\label{beltrami-2''}
\end{equation}
where the harmonic field $\bm{B}_{\rm H} $ is uniquely determined by the fluxes $\Phi_1,\cdots,\Phi_\nu$.
When $\bm{B}_{\rm H} $ and $\mu$ are given, we solve (\ref{beltrami-2''}) for $\bm{B}_\Sigma$
to obtain the Beltrami magnetic field $\bm{B}_\mu=\bm{B}_\Sigma + \bm{B}_{\rm H}$.
If $\bm{B}_{\rm H} =0$, (\ref{beltrami-2''}) has solutions only for discrete eigenvalues $\mu\in\{\lambda_1,\lambda_2,\cdots\} =: \sigma_p(\mathcal{S})$
of the \emph{self-adjoint curl operator} $\mathcal{S}$ defined 
on the operator domain\,\cite{YG1990}
\begin{equation}
D({\cal S}) =  H^1_{\Sigma\Sigma}(\Omega):=\{ \bm{u}\in L^2_{\Sigma}(\Omega)\cap H^1(\Omega);\,
\nabla\times\bm{u}\in L^2_\Sigma(\Omega)\}.
\label{H1SS}
\end{equation}
If $\bm{B}_H\neq0$, 
(\ref{beltrami-2''}) has a nontrivial solution 
for every $\mu\not\in\sigma_p(\mathcal{S})$\,\cite{YG1990}.
Moreover, if the vector potential $\bm{A}_H$ of $\bm{B}_H$ and the eigenfunction $\bm{\omega}_j$
of $\mathcal{S}$ belonging to an eigenvalue $\lambda_j$ are orthogonal (i.e.
$\langle\bm{A}_H,\bm{\omega}_j\rangle=0$, the inhomogeneous equation (\ref{beltrami-2''})
has a solution $\bm{G}_j$ at $\mu=\lambda_j$ even with $\bm{B}_H\neq0$.
Then, $\mu=\lambda_j$ is a \emph{bifurcation point} of two branches of Beltrami fields,
$\bm{B}_\mu$ with $\mu\gtrsim \lambda_j$ and $\bm{B}_{\lambda_j,\alpha} = \bm{G}_j + \alpha\bm{\omega}_j$
($\alpha\in\mathbb{R}$),
and the latter has a smaller energy for a given helicity $C_1$ and $\bm{B}_H$\,\cite{YD2012}.

%---------------------------------------------------------------------------------
\subsubsection{Linearization near the Beltrami equilibrium and tearing mode}

In the neighborhood of a Beltrami equilibrium, we find an infinite number of Casimir elements stemming from the resonant singularity of the Poisson operator, which foliate the phase space and separate the bifurcated Beltrami equilibria on a common helicity leaf.

We linearize the MHD equations.  
Since the Beltrami equilibrium $\bm{u}_\mu=\,^t(0,\bm{B}_\mu)$ is a
stationary point of the energy-Casimir functional $H_\mu = H -\mu C_1$, 
the linearization of Hamilton's equation is rather simple:
Denoting by $\tilde{\bm{u}}=\,^t(\tilde{\bm{V}},\tilde{\bm{B}})$ the perturbed state vector,
we define linearized Hamiltonian  and  Poisson operators  by
\begin{eqnarray}
\mathcal{H}_\mu &=& \left( \begin{array}{cc}
~~1~~ & 0 \\
0 & 1-\mu {\cal S}^{-1}
\end{array} \right),
\label{linear-MHD_Hamiltonian}
\\
{\cal J}_\mu &=& \left( \begin{array}{cc}
0 & \mathcal{P}_\sigma (\textrm{curl} ~\circ) \times\bm{B}_\mu \\
\textrm{curl}( \circ\times\bm{B}_\mu) & 0  
\end{array} \right).
\label{linear-MHD_Poisson}
\end{eqnarray}
Evidently, $\mathcal{H}_\mu$ is a self-adjoint operator for every $\mu\in\mathbb{R}$.
The linearized Hamiltonian equation reads
\begin{equation}
\partial_t \tilde{\bm{u}} = {\cal J}_\mu \mathcal{H}_\mu \tilde{\bm{u}}.
\label{Hamitons_equation-Casimir-linear}
\end{equation} 

In what follows, we assume $\mu>0$.  Then, 
the positive side of the spectrum $\sigma_p({\cal S})$ plays an essential role;
for $\mu<0$, we switch to the negative side of $\sigma_p({\cal S})$.
% Let $\lambda_j$ denote the $j$-th eigenvalue of ${\cal S}$ on the positive side.
Evidently, $\mu\geq\lambda_1$ destroys
the coercivity of $\langle \mathcal{H}_\mu \tilde{\bm{u}},\tilde{\bm{u}}\rangle$
with respect to the norm $\|\tilde{\bm{u}}\|^2$, violating the sufficient condition 
of stability\,\cite{YOIM2003} (see also \cite{lausanne,Holm1985,morrison86,Tasso1992,petrelis}).
In fact, a perturbation $\tilde{\bm{B}}\propto\bm{\omega}_1$ 
(the eigenfunction corresponding to $\lambda_1$) 
yields $\langle \mathcal{H}_\mu \tilde{\bm{u}},\tilde{\bm{u}}\rangle \leq0$. 
However, the negative energy of
a perturbation $\tilde{\bm{B}}\propto\bm{\omega}_1$ does not necessarily cause an ideal-MHD instability, since motion including $\bm{\omega}_1$
may be ``inhibited'' in the Hamiltonian mechanics.

Let us see how Casimir elements foliate  the phase space of perturbations:
$\textrm{Ker}({\cal J}_\mu)$ consists of
two classes of elements:
$\,^t(\bm{v}, 0)$ and $\,^t(0,\bm{b})$ with
$\bm{v}$ and $\bm{b}$ satisfying, respectively, 
\begin{subequations}
% \label{grp}
\begin{align}
 \nabla\times(\bm{B}_\mu\times\bm{v}) &=0,\quad \nabla\cdot\bm{v}=0,
\label{linear-J-v}
\\
 \bm{B}_\mu\times(\nabla\times \bm{b}) &=0.
\label{linear-J-b}
\end{align}
\end{subequations}
The Casimir elements are, in terms of such $\bm{v}$ and $\bm{b}$, 
\begin{equation}
C_{\bm{v}}(\tilde{\bm{u}}) := 
\int \tilde{\bm{V}}\cdot\bm{v}\,\rmd^3x,
\quad
C_{\bm{b}}(\tilde{\bm{u}}) := 
\int \tilde{\bm{B}}\cdot\bm{b}\,\rmd^3x.
\label{linear_Casimir-b} 
\end{equation}
Obviously, we can choose $\bm{v}=\bm{B}_\mu$ and $\bm{b}=\bm{B}_\mu$.
However, far richer solutions stem from the \emph{singularity} of ${\cal J}_\mu$.

Here we concentrate on the ``magnetic part'' (\ref{linear-J-b}),
but a similar singular solution $\bm{v}$ can be constructed for 
the ``flow part'' (\ref{linear-J-v}).
The determining equation (\ref{linear-J-b}) of $\bm{b}$ can be rewritten as
\begin{equation}
\nabla\times\bm{b}= \eta\bm{B}_\mu
\label{linear-J-b'}
\end{equation}
with some scalar function $\eta$.  
We have already found a solution $\bm{b}=\bm{B}_\mu$ and $\eta=\mu$.
Here we seek solutions with non-constant $\eta$.
However, $\eta$ is not a free function;
the divergence of both sides of (\ref{linear-J-b'}) yields
\begin{equation}
\bm{B}_\mu\cdot\nabla\eta =0,
\label{linear-J-b-integrability}
\end{equation}
which implies that $\eta$ is constant along the magnetic field lines.
For the integrability of $\eta$, the equilibrium field $\bm{B}_\mu$ must have 
integrable field lines; a continuous spatial symmetry guarantees this.
Here we consider a \emph{slab geometry}, in which we may write
$\bm{B}_\mu = \,^t\left( 0 , B_y(x) , B_z(x) \right)$.
% Each surface of $x=$ constant is a magnetic surface (integral surface of field lines).
Denoting $\bm{b}=  \,^t\left( 0 , b_y(x), b_z(x) \right)$,
(\ref{linear-J-b}) reads as
\begin{equation}
B_y\partial_x b_y + B_z\partial_x b_z =0,
\label{slab_Casimir2}
\end{equation}
which may be solved for $b_y(x)$, given an arbitrary $b_z(x)$.
Furthermore, we have \emph{singular} (hyper-function) solutions; let us consider
\begin{equation}
\bm{b}=  \,^t\left( 0,  b_y(x), b_z(x)\right)  \rme^{\rmi(k_yy+k_zz)}.
\label{linear-J-b0}
\end{equation}
Putting $b_y(x)=\rmi k_y\vartheta(x)$ and $b_z(x)=\rmi k_z\vartheta(x)$, (\ref{slab_Casimir2}) reduces into
\begin{equation}
[B_y(x)k_y + B_z(x)k_z ]\partial_x \vartheta(x) = 0,
\label{linear-J-b1}
\end{equation}
% and $b_z(x)=(k_z/k_y)b_y(x)$, (\ref{linear-J-b}) holds
% (here we assume $k_y\neq0$; otherwise, we put $b_y=0$, and replace $b_y$ by $b_z$ in
% (\ref{linear-J-b1})).
which yields
\begin{equation}
\vartheta(x) = c_0 + c_1 Y(x-x^\dagger),
\label{linear-J-b2}
\end{equation}
where % $Y(\cdot)$ is the  Heaviside step function, 
$c_0, \, c_1$ are complex constants, and
$k_y$, $k_z$ and $x^\dagger$ (real constants) are chosen to
satisfy the \emph{resonance condition}
\begin{equation}
 B_y(x^\dagger)k_y + B_z(x^\dagger)k_z =0 .
\label{B-resonance}
\end{equation}
Then, 
$\eta = \rmi (k_y/B_z) \rme^{\rmi(k_yy+k_zz)}\delta(x-x^\dagger)$.
% = \frac{-\rmi k_z  \rme^{\rmi(k_yy+k_zz)}}{B_y}\delta(x-x^\dagger),
% \label{slab_Casimir3}
% \end{equation}
% which formally satisfies (\ref{linear-J-b-integrability}).
>From (\ref{linear-J-b'}) we see that this Dirac $\delta$-function solution implies a \emph{current sheet} on the resonant surface $\Gamma^\dagger: x=x^\dagger$. 
Physically, $\Gamma^\dagger$ represents a thin layer of ideal-MHD plasma that supports a sheet current, which suppresses the change of field-line topology\, \cite{Boozer_Pomphrey_11}.

%---------------------------------------------------------------------------------
% \subsubsection{Helical-flux Casimir invariants}
In what follows, we normalize the kernel element $\bm{b}$ so that $\|\bm{b}\|^2
= \langle \bm{b},\bm{b}\rangle = 1$.
The singular (hyper-function) solution $\bm{b}$ of (\ref{linear-J-b0})
created by the resonance singularity (\ref{B-resonance}), imposes an essential restriction on the
range of dynamics;
any magnetic perturbation $\tilde{\bm{B}}$ such that $\langle\tilde{\bm{B}},\bm{b}\rangle\neq0$
is forbidden to change, because 
\begin{equation}
C_{\bm{b}}(\tilde{\bm{u}})=C_{\bm{b}}(\tilde{\bm{B}}):=\langle \tilde{\bm{B}},\bm{b}\rangle
\label{helical-flux_Casimir}
\end{equation}
is an invariant. 
We call $C_{\bm{b}}(\tilde{\bm{B}})$ a ``helical-flux Casimir invariant.''
The equilibrium point of the energy-Casimir functional 
\begin{equation}
{\cal F}_{\mu,\beta}(\tilde{\bm{u}}):=
\frac{1}{2}\langle \mathcal{H}_\mu \tilde{\bm{u}},\tilde{\bm{u}}\rangle  -
% \sum_{\bm{b}} 
\beta C_{\bm{b}}(\tilde{\bm{u}})
\label{linear_energy-Casimir}
\end{equation}
gives the \emph{tearing mode}.

Because of the linearity of the determining equation
(\ref{linear-J-b0}),
the totality of $\,^t(0,\bm{b})\in\textrm{Ker}({\cal J}_\mu)$ is a linear subspace 
of the total phase space and it is ``integrable'' -- thus  foliates the phase space in terms of
the Casimir invariants $C_{\bm{b}}(\tilde{\bm{u}})=\langle \tilde{\bm{B}},\bm{b}\rangle$.
In the next subsection, we choose the ``dominant helical-flux Casimir''
that has the common Fourier coefficients with the helical mode $\bm{\omega}_1$
of the bifurcated helical Beltrami equilibrium, and define the
``minimum extension'' that canonizes the corresponding kernel of ${\cal J}_\mu$.

%%%%%%%%%%%%%%%%%%%%%%%%%%
%%%%%%%%%%%%%%%%%%%%%%%%%%%%%%%%%%%%
%%%%%%%%%%%%%%%%%%%%%%%%%%%
\subsection{Tearing-mode instability}
\label{subsec:tearing-mode}

%---------------------------------------------------------------------------------
\subsubsection{Canonization}

Let $\,^t(0,\bm{b})\in\textrm{Ker}({\cal J}_\mu)$.
We separate a one-dimensional subspace $\{p\bm{b};\, p\in\mathbb{R}\}$ from the
phase space $L^2_\Sigma(\Omega)$ of magnetic perturbations $\tilde{\bm{B}}$,
and denote by $\wp_\parallel$ the orthogonal projection onto the remaining space:
% (the leaf foliated by the Casimir invariant $C_{\bm{b}}(\tilde{\bm{u}})$),
\[
\wp_\parallel \tilde{\bm{B}} := \tilde{\bm{B}} - \langle \tilde{\bm{B}} ,\bm{b} \rangle \bm{b} .
\]
We also denote 
\[
\wp_\perp \tilde{\bm{B}} := \langle \tilde{\bm{B}} ,\bm{b} \rangle \bm{b} = C_{\bm{b}}(\tilde{\bm{B}})\bm{b},
\]
and decompose
$\tilde{\bm{B}} = \wp_\parallel \tilde{\bm{B}} + \wp_\perp\tilde{\bm{B}}$.
% and write $p:= \langle \tilde{\bm{B}} ,\bm{b} \rangle = C_{\bm{b}}(\tilde{\bm{u}})$.
Writing the state vector as
$\tilde{\bm{u}}' = \,^t(\tilde{\bm{V}}, \wp_\parallel\tilde{\bm{B}},\wp_\perp\tilde{\bm{B}})$,
and denoting ${\cal K}_\mu:=(1-\mu {\cal S}^{-1})$,
the Hamiltonian and Poisson operators read
% denoting
% \begin{equation}
% \alpha:=\langle (1-\mu {\cal S}^{-1})\bm{b},\bm{b}\rangle ,
% \label{alpha}
% \end{equation}
\begin{eqnarray}
\mathcal{H}_\mu' &=& \left( \begin{array}{cc:c}
 ~~1~~ & 0 & ~\\
     0 & \wp_\parallel{\cal K}_\mu & \wp_\parallel{\cal K}_\mu \\
\hdashline
~& \wp_\perp{\cal K}_\mu & \wp_\perp{\cal K}_\mu
\end{array} \right),
\label{H_mu-2}
\\
{\cal J}_\mu' &=&
\left( \begin{array}{cc:c}
% 0  & -\nabla\cdot & 0 & ~ \\
 0 & (\mathrm{curl} \wp_\parallel~\circ) \times\bm{B}_\mu & ~ \\
 \wp_\parallel\mathrm{curl}( \circ\times\bm{B}_\mu) & 0  & ~ \\
\hdashline
~ & ~ & 0
\end{array} \right).
\label{linear-J-2}
\end{eqnarray}
Notice that the kernel $^t(0,\bm{b})$ has been separated from the upper left block of 
the Poisson operator. 
% Whole other elements are zero, since $\wp_\perp\tilde{\bm{B}} \in \textrm{Ker}({\cal J}')=\textrm{Coker}({\cal J}')$.

Now, we introduce an adjoint variable $q$ to extend the phase space:
\[
\tilde{\bm{u}}_{ex} 
= ~^t(\tilde{\bm{V}}, \wp_\parallel\tilde{\bm{B}}, \wp_\parallel\tilde{\bm{B}},q)
\]
and define 
\begin{equation}
{\cal J}_{\mu,ex} =
\left( \begin{array}{cc:cc}
 0 & (\mathrm{curl} \wp_\parallel~\circ) \times\bm{B}_\mu & ~& ~ \\
\wp_\parallel\mathrm{curl}( \circ\times\bm{B}_\mu) & 0  & ~ & ~ \\
\hdashline
 ~ & ~ & 0 & -1 \\
 ~ & ~ &1 & 0 
\end{array} \right),
\label{linear-J-3}
\end{equation}
which is ``canonized'' by extending  the variable $p$.
Since the original system does not include $q$ as an variable, we may write
\begin{equation}
\mathcal{H}_{\mu,ex} = \left( \begin{array}{cc:cc}
 ~~1~~ & 0 & ~& ~\\
 0 & \wp_\parallel{\cal K}_\mu & \wp_\parallel{\cal K}_\mu & ~\\
\hdashline
 ~& \wp_\perp{\cal K}_\mu & \wp_\perp{\cal K}_\mu & 0 \\
 ~& ~        & 0 &  0
\end{array} \right).
\label{H_mu-2'}
\end{equation}
Evidently, $\wp_\perp\tilde{\bm{B}}= \langle \tilde{\bm{B}} ,\bm{b} \rangle \bm{b}$
is invariant, which is originally a Casimir element, but
is now an invariant due to the symmetry of $\mathcal{H}_{\mu,ex}$ with respect to the
new variable $q$.

%---------------------------------------------------------------------------------
\subsubsection{Singular perturbation}

Perturbing the Hamiltonian with respect to $q$,
we can break the invariance of $p:=\langle \tilde{\bm{B}} ,\bm{b} \rangle=C_{\bm{b}}(\tilde{\bm{B}})$.
We consider a Hamiltonian
\begin{equation}
\mathcal{H}_{\mu,EX} := \left( \begin{array}{cc:cc}
~~1~~ & 0 & ~& ~\\
0 & \wp_\parallel{\cal K}_\mu & \wp_\parallel{\cal K}_\mu & ~\\
\hdashline
~& \wp_\perp{\cal K}_\mu & \wp_\perp{\cal K}_\mu & 0 \\
~& ~        & 0 & D
\end{array} \right),
\label{H_mu-3}
\end{equation}
where $D$ is a parameter that is  introduced to couple  the
original system to the  \emph{external variable} $q$.
Note that the original energy $\langle \mathcal{H}_\mu \tilde{\bm{u}},\tilde{\bm{u}}\rangle/2$
is no longer an invariant; instead, the new total energy
$\langle \mathcal{H}_{\mu,EX} \tilde{\bm{u}}_{ex},\tilde{\bm{u}}_{ex} \rangle/2$ is conserved.

The induced change in the Casimir element (helical flux, which is now 
denoted by $p$) is estimated by the canonized block of   Hamilton's equations:
\begin{equation}
\frac{\rmd}{\rmd t} p = -D q, \quad
\frac{\rmd}{\rmd t} q = \langle {\cal K}_\mu \tilde{\bm{B}}, \bm{b}\rangle .
\label{Casimir-perturbation-1}
\end{equation}
% which combines as (assuming that $D$ is temporally constant)
% \begin{equation}
% \frac{d^2}{dt^2} p = - D \langle {\cal K}_\mu \tilde{\bm{B}}, \bm{b}\rangle .
% \label{Casimir-perturbation-2}
% \end{equation}
% Evidently, $X\neq0$ breaks the constancy of the helical flux Casimir invariant $p$.
For $\tilde{\bm{B}}=p\bm{\omega}_1$ (the eigenfunction determining the bifurcated fiducial-energy
equilibrium), we may estimate
\[
\langle {\cal K}_\mu \tilde{\bm{B}}, \bm{b}\rangle 
=  \langle (1-\mu{\cal S}^{-1}) \bm{\omega}_1, \bm{b}\rangle p
= (1-\mu/\lambda_1)\langle \bm{\omega}_1, \bm{b}\rangle p .
\]
Absorbing the sign of $\langle \bm{\omega}_1, \bm{b}\rangle$ by $p$,
we assume $\gamma := \langle \bm{\omega}_1, \bm{b}\rangle >0$.
For simplicity, let us assume that $D$ is a constant number.
The system (\ref{Casimir-perturbation-1}) has the  Hamiltonian
\begin{equation}
H_p :=  \left(1-\frac{\mu}{\lambda_1} \right) \gamma \frac{ p^2}{2} + D \frac{q^2}{2} .
\label{Casimir-Hamiltonian}
\end{equation}
This sub-system Hamiltonian describes the coupling of the original (un-perturbed)
Hamiltonian system with an ``external energy'' $Dq^2/2$.
If this external energy is positive (i.e. $D>0$), 
the ``internal energy'' of the original
system may ``dissipate'' through the coupling.
% ; hence assuming $D>0$ is
% physically appropriate for the present purpose of modeling a ``dissipative
% breaking'' of the Casimir foliation.
The factor
$(1-\mu/\lambda_1)$ of  the ``kinetic energy'' part of the Hamiltonian $H_p$
may be interpreted as an effective (reciprocal) mass of the tearing mode
% (the ideal Casimir representing the current sheet at the resonant surface)
---beyond the \emph{bifurcation point} $\mu=\lambda_1$, the effective mass
becomes negative, and the ``negative-energy mode'' can grow by absorbing
energy from the positive energy source $Dq^2/2$.

%%%%%%%%%%%%%%%%%%%%%%%%%%%%%%%%%%%%%%%%%%%%
%%%%%%%%%%%%%%%%%%%%%%%%%%%%%%%%%%%%%%%%%%%%%%%%%%
%%%%%%%%%%%%%%%%%%%%%%%%%%%%%%%%%%%%%%%%%%
%%%%%%%%%%%%%%%%%%%%%%%%%%%%%%%%%%%%%%%%%%%%%%%%%%%%
%%%%%%%%%%%%%%%%%%%%%%%%%%%%%%%%%%%%%%%%%
%%%%%%%%%%%%%%%%%%%%%%%%%%%%%%%%%%%%%%%%%%%%%%%%%%%%%
\section{Conclusion}
\label{sec:conclusion}

We have described several facets of noncanonical Hamiltonian systems.  Namely, 
that  the Poisson operator (field tensor)
of a noncanonical Hamiltonian system has a nontrivial kernel 
(and thus, a cokernel) that foliates the phase space (Poisson manifold),
imposing topological constraints on the dynamics.
The Hamiltonian (energy) of a weakly-coupled macroscopic system (such as a 
normal fluid or a plasma) is usually rather simple 
---it being a convex functional (typically a quadratic form) by which
one can define an energy norm on the phase space.
However, an ``effective energy'' may have a considerably nontrivial distribution on 
the actual phase space of constrained variables,
which is a ``distorted'' manifold (or a leaf) immersed in the total space.
Interesting structures created in a fluid or a plasma may be delineated by 
unearthing leaves of the phase space
and analyzing their distortion with respect to the energy norm.
When one can ``integrate'' the kernel of the Poisson operator to construct Casimir elements,
the Casimir leafs foliate the Poisson manifold and, then, the effective energy is
the energy-Casimir functional.

In addition, here we have proposed a model for a  physical process that removes the constraints of Casimir elements
and enables the system to seek  lower-energy states on different Casimir leaves.
By invoking an extended phase-space, we  canonized the Poisson operator and introduced  
a coupling of the original ideal system with an external energy source
---the exchange of energy between the original system and the connected external system
thus described a  ``dissipation'' process.
This formulation is based on the method of ``minimum canonization'' 
that interprets Casimir elements as ``adiabatic invariants,'' and
``unfreezes'' the Casimir elements to be dynamic,  by perturbing 
the Hamiltonian with respect to the new angle variable added to the phase space;
such perturbations  increase the number of degrees of freedom and are therefore a  kind of \emph{singular perturbation}.

The theory was applied to the tearing-mode instability, where a 
 tearing mode was  regarded as an equilibrium point on a helical-flux Casimir leaf.
As long as the helical-flux is constrained, the tearing mode cannot grow. However,  it was shown that 
a singular perturbation that allows the system to change the
helical flux can cause a tearing mode to grow if it has an excess energy with respect to a
fiducial energy of the Beltrami equilibrium at the bifurcation point.

\bigskip
\textbf{Acknowledgements.}
The authors acknowledge  discussions with and suggestions of
 S. M. Mahajan,  R. L. Dewar, and  F.  Dobarro. 
ZY was supported by the Grant-in-Aid for Scientific Research 
No. 23224014 from MEXT-Japan.
PJM was supported by US Department of Energy, Grant No.~DE-FG02-04ER54742.

 %%%%%%%%%%%%%%
\bibliographystyle{ouvrage-hermes}

\begin{thebibliography}{ }

\bibitem{Arnold-Khesin}
Arnold V. I., Khesin B.A., 
{\it Topological Methods in Hydrodynamics},
(Springer, 1998).

\bibitem{Boozer_Pomphrey_11}
Boozer A. H., Pomphrey N.,
Current density and plasma displacement near perturbed rational surfaces
\textit{Phys. Plasmas} \textbf{17} (2011) 110707-1--4.

\bibitem{chandre}
Chandre C, Morrison P. J.,  Tassi E.,  On the Hamiltonian Formulation of Incompressible Ideal Fluids and Magnetohydrodynamics via Dirac's Theory of Constraints,
\textit{Physics Letters A} \textbf{376}  (2012) 737--743.

\bibitem{Clarke1975}
Clarke F. H.,
Generalized gradients and applications,
\textit{Trans Amer Math Soc} \textbf{205} (1975) 247--262.

\bibitem{FKR63}
Furth H. P., Killeen J., Rosenbluth M. N.,
Finite Resistivity instabilities of a sheet pinch,
\textit{Phys. Fluids} \textbf{6} (1963) 459--484.

\bibitem{Furth63}
Furth H. P.,
Hydromagnetic instabilities due to finite resistivity,
in \textit{Propagation and Instabilities in Plasmas}, Ed. Futterman W. T.
(Stanford Univ. Press, 1963) PP. 87--102.

\bibitem{Furth63}
Furth H. P.,
Hydromagnetic instabilities due to finite resistivity,
in \textit{Propagation and Instabilities in Plasmas}, Ed. Futterman W. T.
(Stanford Univ. Press, 1963) PP. 87--102.

\bibitem{Holm1985}
Holm D. D., Marsden J. E., Ratiu T., Weinstein A.,
Nonlinear stability of fluid and plasma equilibria,
\textit{Phys. Rep.} \textbf{123} (1985) 1--116. 

\bibitem{lausanne}
Hazeltine R. D., Holm D. D., Marsden J. E.,  Morrison P. J., 
Generalized Poisson Brackets and Nonlinear Liapunov Stability--Application to Reduced MHD,
in \textit{International Conference on Plasma Physics Proceedings 1}, 
Eds. M.Q. Tran and M.L. Sawley, (Ecole Polytechnique Federale de Lausanne, Lausanne, 1984) P. 203.

\bibitem{khesin} 
Khesin B., Wendt R., 
{\it The Geometry of Infinite-Dimensional Groups},
(Springer-Verlag, 2009).

\bibitem{energy-casimir}  
Kruskal M. D., Oberman C.,
On the stability of plasma in static equilibrium,
\textit{Phys. Fluids} \textbf{1} (1958), 275--280.

\bibitem{Moffatt}
Moffatt H. K.,
\textit{Magnetic field generation in electrically conducting fluids},
(Cambridge University Press. 1978).

\bibitem{MG80}  
Morrison P. J., Greene J. M.,
Noncanonical Hamiltonian density formulation of hydrodynamics 
and ideal magnetohydrodynamics,
\textit{Phys. Rev. Lett.} \textbf{45} (1980), 790--794.
 

\bibitem{morrison86}
Morrison  P. J. ,   Eliezer S., 
Spontaneous Symmetry Breaking and Neutral Stability in the Noncanonical {H}amiltonian Formalism, 
\textit{Phys. Rev. A} \textbf{33}  (1986) 4205--4214.


\bibitem{morrison87}
Morrison  P. J.,
Variational Principle and Stability of Nonmonotonic Vlasov-Poisson Equilibria,
\textit{Zeitschrift f\"ur Naturforschung} \textbf{42a}  (1987) 1115--1123.

\bibitem{morrison98} 
Morrison P. J.,
Hamiltonian description of the ideal fluid,
\textit{Rev. Mod. Phys.} \textbf{70} (1998) 467--521.

\bibitem{narayanan} 
Narayanan V., Morrison P. J.,
Rank change in Poisson dynamical systems,
\textit{arXiv:1302.7267v1 [math-ph] 28 Feb 2013.}

\bibitem{petrelis}
P\'etr\'elis F., Alexakis A., Doering  C. R.,  Morrison  P. J.,
Bounds on Dissipation in Magnetohydrodynamic Problems in Plane Shear Geometry,
\textit{Phys. Plasmas} \textbf{10} (2003) 4314--4323.

\bibitem{Tasso1992}
Tasso H.,
Simplifies version of a stability condition in resistive MHD,
\textit{Phys. Lett. A.} \textbf{169} (1992) 396--398.

\bibitem{JBT74}
Taylor J. B.,
Relaxation of toroidal plasma and generation of reverse magnetic fields,
\textit{Phys. Rev. Lett,} \textbf{33} (1974) 1139--1141.

\bibitem{JBT86}
Taylor J. B.,
Relaxation  and magnetic reconnection in plasmas.
\textit{Rev. Mod. Phys.} \textbf{58} (1986) 741--763.

\bibitem{White83}
White R. B.,
Resistive instabilities and field-line reconnection,
in \textit{Basic Plasma Physics} Vol. 1, Ed. Galeev A A and Sudan R N
(North-Holland, 1983) PP. 611--676.

\bibitem{YG1990}
Yoshida Z,, Giga Y.,
Remarks on spectra of operator rot,
\textit{Math. Z.} \textbf{204} (1990) 235--245.

\bibitem{YOIM2003}
Yoshida Z., Ohsaki S., Ito A., Mahajan S. M.,
Stability of Beltrami flows,
\textit{J. Math. Phys.} \textbf{44} (2003) 2168--2178.

\bibitem{Yoshida_Springer}
Yoshida Z.,
\textit{Nonlinear Science ---The Challenge of Complex Systems}, 
(Springer-Verlag, 2010).

\bibitem{YMD}
Yoshida Z., Morrison P.J., Dobarro F.,
Singular Casimir elements of the Euler equation and equilibrium points,
\textit{J. Math. Fluid Mech.}  accepted (2013). 
\textit{arXiv:1107.5118 [math-ph] 30 Jul 2011}.

\bibitem{YD2012}
Yoshida Z., Dewar R. L.,
Helical bifurcation and tearing mode in a plasma ---a description based on Casimir foliation,
\textit{J. Phys. A: Math. Theor.} 
\textbf{45} (2012) 365502 1--36.

\bibitem{RT-1_PPCF2013}
Yoshida Z., Saitoh H., Yano Y., Mikami H., Kasaoka N., Sakamoto W., Morikawa J., Furukawa M.,
Mahajan S. M.,
Self-organized confinement by magnetic dipole: recent results from RT-1 and theoretical modeling,
\textit{Plasma Phys. Control. Fusion}
\textbf{55} (2013) 014018 1--5.

 

%%%%%%%%%%%%%%%%%%


\end{thebibliography}

 \end{document}